\documentclass[aps,prc,twocolumn,nofootinbib]{revtex4}
\usepackage[hyperindex=true]{hyperref}
\usepackage[utf8x]{inputenc}
\usepackage{yfonts,tikz}
\usepackage[version=3,arrows=pgf]{mhchem}
\usepackage{amssymb,amsmath}
\usepackage{color}

\newcommand{\bra}[1]{\langle #1 \vert}
\newcommand{\ket}[1]{\vert #1 \rangle}
\newcommand{\bran}[1]{\langle #1}
\newcommand{\vcrea}{\hat{\boldsymbol{a}}^\dagger}
\newcommand{\vanni}{\hat{\boldsymbol{a}}}
\newcommand{\vbcrea}{\hat{\boldsymbol{b}}^\dagger}
\newcommand{\vbanni}{\hat{\boldsymbol{b}}}
\newcommand{\vccrea}{\hat{\boldsymbol{c}}^\dagger}
\newcommand{\vcanni}{\hat{\boldsymbol{c}}}

\newcommand{\anni}{\hat{a}}
\newcommand{\bcrea}{\hat{b}^\dagger}
\newcommand{\banni}{\hat{b}}
\newcommand{\ccrea}{\hat{c}^\dagger}
\newcommand{\canni}{\hat{c}}
\newcommand{\vgvb}{\boldsymbol{\bar{z}}}
\newcommand{\vgv}{\boldsymbol{z}}
\newcommand{\vgvbe}{\boldsymbol{\bar{\eta}}}
\newcommand{\vgve}{\boldsymbol{\eta}}

\newcommand{\vgvz}{\boldsymbol{\zeta}}

\newcommand{\gvb}{\bar{z}}
\newcommand{\gv}{z}
\newcommand{\gvbe}{\bar{\eta}}
\newcommand{\gve}{\eta} 
\newcommand{\gvz}{\zeta}
\newcommand{\Id}{\openone}

\newcommand{\dnl}{\,\mathrm{d}^{n}_{\tiny \triangleleft}}
\newcommand{\dnr}{\,\mathrm{d}^{n}_{\tiny \triangleright}}

\newcommand{\Gvar}{$\mathcal{G}$-variables}

\newcommand\ncoord[2][0,0]{\tikz[remember picture,overlay]{\path (#1) coordinate (#2);}}

\newcommand{\nuc}[2]{$^{#1}${#2}}

\begin{document}

\title{Evaluation of overlaps between arbitrary Fermionic quasiparticle vacua}

\author{B. Avez}
\affiliation{Universit\'e Bordeaux 1, CNRS/IN2P3, Centre d'\'Etudes Nucl\'eaires de Bordeaux Gradignan, 
             Chemin du Solarium, BP120, 33175 Gradignan, France}

\author{M. Bender}
\affiliation{Universit\'e Bordeaux 1, CNRS/IN2P3, Centre d'\'Etudes Nucl\'eaires de Bordeaux Gradignan, 
             Chemin du Solarium, BP120, 33175 Gradignan, France}

\date{\today}

\begin{abstract}
We derive an expression that allows for the unambiguous evaluation 
of the overlap between two arbitrary quasiparticle vacua, including 
its sign. Our expression is based on the Pfaffian of a skew-symmetric 
matrix, extending the formula recently proposed by
[L.~M.\ Robledo, Phys.\ Rev.\ C \textbf{79}, 021302(R) (2009)] to 
the most general case, including the one of the overlap between two 
different blocked $n$-quasiparticle states for either even or odd systems.  
The powerfulness of the method is illustrated for a few typical matrix
elements that appear in realistic angular-momentum-restored
Generator-Coordinate Method calculations when breaking time-reversal
invariance and using the full model space of occupied single-particle 
states.
\end{abstract}


\maketitle

\section{Introduction}

The evaluation of kernels in projection and more general applications 
of the Generator Coordinate Method (GCM) based
on quasiparticle vacua requires the calculation of the overlap
between two different quasiparticle vacua. 
Its evaluation presents 
a long-standing technical challenge: the standard expression
for this overlap, the so-called "Onishi formula" \cite{Oni66} 
provides the square of the (complex) overlap only. As a consequence, 
the overall sign of the overlap is not determined, or, equivalently, 
its phase is determined up to integer multiples of $\pi$ only.

One possible solution to the problem was proposed by Neerg{\aa}rd and
W{\"u}st~\cite{Nee83}, but the practical application of their 
technique becomes cumbersome in realistic applications and has been
rarely used in practice.
Notable exceptions are Refs.~\cite{Schm04,Rod10}.
Many groups have resided to determine the phase through a kind of 
Taylor expansion that allows "to follow the overlap" when the kernel 
can be connected in small steps to a known reference overlap (see, 
for example, Refs.~\cite{Har82,Ena99}). In practice this can become 
very cumbersome or even impossible when the phase is rapidly changing 
or when there is no symmetry that establishes a reference phase. 

It was pointed out by Robledo in Ref.~\cite{Rob09} that techniques 
for the manipulation of matrix elements between Fermionic coherent 
states that are well-known in field theory allow to express the 
overlap between two quasiparticle vacua, including its sign, as 
the so-called Pfaffian of a skew-symmetric matrix.\footnote{
The possibility to use Pfaffians for this purpose was already conjectured 
much earlier by Balian and Br{\'e}zin in Ref.~\cite{Bal69},
but never worked out.}
In a more recent paper, Robledo~\cite{Rob11} has also worked out 
the practical implementation of this idea for the unambiguous 
evaluation of the overlap between fully-paired quasiparticle vacua,
including the limit where some of the "pairs" consist of fully
occupied single-particle states.
Even more recently, Bertsch and Robledo~\cite{Ber11x} also 
investigated the case of systems with an odd number of constituants, 
providing the unambiguous evalution of the overlap for the
special case of two quasiparticle vacua linked by a symmetry 
transformation.

In the present paper, we present the extension of this scheme to
the calculation of the overlap between two possibly different
arbitrary quasiparticle vacua, generalizing the treatment of
completely filled single-particle states to the most general case.   
This extension thus allows to handle quasiparticle vacua obtained
from blocked 1-, 2-, \ldots $n$ arbitrary quasiparticle states.   
To this aim, we present an alternative derivation of the overlap 
that makes use of an extension of the standard Thouless parameterization 
of quasiparticle vacua~\cite{Tho60} that is advantageous in the presence 
of completely filled single-particle states and allows to avoid many of 
the matrix manipulations elaborated in Ref.~\cite{Rob11}. Also, our 
final expression Eq.~(\ref{eq:result}) allows for the calculation 
of the overlap of two quasiparticle vacua that are expressed in different 
single-particle bases that do not span the same sub-space of the Hilbert 
space of single-particle states, a situation frequently encountered in 
symmetry-restored GCM codes that use a coordinate space representation 
of the quasiparticle vacua in terms of their canonical single-particle 
bases~\cite{Flo75,Bon90,Val00,Ben09a,Ben11x}.

The article is organized as follows: Section~II introduces a 
generalization of the Thouless parameterization of quasiparticle 
vacua for blocked states that will turn out to be useful for the 
purpose of our paper. 
Section~III reviews key properties of Fermionic coherent states
and Grassmann calculus that will be needed lateron and thereby 
introduces our notation.
Section~IV describes the calculation of the overlap, and 
Section~V presents some illustrative examples of overlaps 
from realistic calculations. Finally,  
Section~VI summarizes our findings. 
An appendix gives the representation of determinants and Pfaffians 
of skew-symmetric matrices in terms of integrals over Grassmann 
variables.

\section{Parameterization of quasiparticle vacua}

Let $\{\vcrea,\vanni\}$ and $\{\vbcrea,\vbanni\}$ be two not necessarily equivalent single-particle 
bases of dimension $n$ ($n$ \emph{even}), with which are defined two (not normalized) quasiparticle 
vacua $\ket{\phi_a}$ and $\ket{\phi_b}$ through the parameterizations\footnote{
For time being, $n$ will be the minimal number of single-particle states that allows to represent 
both $\ket{\phi_a}$ and $\ket{\phi_b}$ in their respective single-particle basis.
}
\begin{equation}
\ket{\phi_{c}} 
= e^{\frac{1}{2}\sum_{kl}M^{(c)}_{kl}\ccrea_k\ccrea_l} \prod^{\tiny \triangleright}_{i\in\{I^{c}\}}\ccrea_{i}\ket{-} 
\, ,
\label{eq:VacParm}
\end{equation}
where $c$ is either $a$ or $b$, $\ket{-}$ is the bare vacuum of single-particle operators,
and where the $M^{(c)}$ are skew-symmetric matrices. Their 
elements with indices belonging to the ensemble of fully occupied states $\{I^{c}\}$ of 
cardinality\footnote{The cardinality is the number of elements of an ensemble.} 
$\#I^c=r_c$ are (can be) put
to zero. This constitutes a natural way to regularize the matrix $M^{(c)}$ in the presence 
of fully-occupied states.

The triangle pointing to the right on top of the product sign means that it is a "direct" product, 
\begin{eqnarray}
\prod^{\tiny \triangleright}_{i\in\{I^{c}\}}\ccrea_{i} = \ccrea_{\mu}\cdots\ccrea_{\nu}\ccrea_{\delta}, 
\quad \mbox{with} \quad \mu>\cdots>\nu>\delta \, ,
\label{eq:directproduct}
\end{eqnarray}
as opposed to a "reverse" product obtained, for example, by taking the adjoint of Eq.~(\ref{eq:directproduct}), i.e.
\begin{eqnarray}
\prod^{\tiny \triangleleft}_{i\in\{I^{c}\}}\canni_{i} = \canni_{\delta}\canni_{\nu}\cdots\canni_{\mu}, 
\quad \mbox{with} \quad \mu>\cdots>\nu>\delta \, .
\label{eq:reverseproduct}
\end{eqnarray}
The parameterization Eq.~(\ref{eq:VacParm}) of quasiparticle vacua is not very different from the one 
by Thouless~\cite{Tho60}.
However, it has two advantages important for our purpose. 
First, this parameterization is well-defined when dealing with fully occupied states.
And second, it allows to parameterize a quasiparticle vacuum 
for systems with odd particle number in terms of single-particle states, something that cannot be achieved 
with the standard Thouless formula. However, this parameterization can be set-up only in a specific
single-particle basis that separates the fully-occupied single-particle states from the others.
Such a single-particle basis is, for example, the canonical single-particle basis of a quasiparticle 
vacuum. The use of these bases does not impose a serious restriction, since they provide the most 
compact representation of a quasiparticle vacuum, such that their use is often desirable in numerical 
applications.

Finally, the convention for the ordering of single-particle levels in Eq.~(\ref{eq:VacParm}) (matrix elements 
of $M^{(c)}$ and product ordering) is to be kept unchanged for \emph{each} calculation involving a given 
state $\ket{\phi_{c}}$. In fact, Eq.~(\ref{eq:VacParm}) is just an alternative convention that circumvents 
the convention of Ref.~\cite{Rob09} to connect the phase of $\langle \phi_a  | \phi_b \rangle$ to 
$\langle \phi_a  | - \rangle$ and $\langle - | \phi_b \rangle$, which cannot be achieved when either 
$\langle \phi_a  | - \rangle$ or $\langle - | \phi_b \rangle$ (or both) is (are) zero.

\section{A reminder on Fermionic coherent states and Grassmann calculus}

In order to evaluate the overlap $\bran{\phi_a}\ket{\phi_b}$ between the two states,
we introduce, following closely Ref.~\cite{Rob09}, two sets of Fermionic coherent states
\begin{eqnarray}
\ket{\vgv_{c}}&=&e^{\vccrea.\vgv_c}\ket{-} \label{eq:FCSr}\\
\bra{\vgv_{c}}&=&\bra{-}e^{\vgvb_c.\vcanni}\label{eq:FCSl}  
\end{eqnarray}
for $c=a,b$, parameterized in terms of anticommuting $\gv_{c_k}$ and $\gvb_{c_k}$ elements 
of a Grassmann algebra $\mathcal{G}$, where the notations $\vccrea.\vgv_c$ and $\vgvb_c.\vcanni$ 
used in Eq.~(\ref{eq:FCSr}) and Eq.~(\ref{eq:FCSl}) stand for
\begin{eqnarray}
\vccrea.\vgv_c  &=& \sum_{i=1}^n \ccrea_i \gv_{c_i}, \\
\vgvb_c.\vcanni &=& \sum_{i=1}^n \gvb_{c_i}\canni_i.
\end{eqnarray}
In particular, we notice that the coherent states $\ket{\vgv_{c}}$ are not normalized. 
Instead, one has $\bran{-}\ket{\vgv_{c}}=\bran{\vgv_{c}}\ket{-}=1$.

In what follows, we recall some useful properties of Grassmann algebra, its associated calculus, 
and of Fermionic coherent states that will be needed for the formal derivations outlined below.
Concerning Grassmann algebra and calculus~\cite{Ber66,Bla85,Zin02}, we recall that
\begin{itemize}
\item The adjoint operator performs a one-to-one mapping within $\mathcal{G}$
\begin{eqnarray}
\left(\gv_{c_k}\right)^\dagger  &=& \gvb_{c_k}, \\
\left(\gvb_{c_k}\right)^\dagger &=& \gv_{c_k},  \\
\left(\gv_{c_k}\gv_{c_l}\right)^\dagger &=& \gvb_{c_l}\gvb_{c_k}.
\end{eqnarray}
\item Grassmann variables anticommute
\begin{eqnarray}
 \gv_{c_k} \gv_{c_l}   &=& - \gv_{c_l} \gv_{c_k},  \\
 \gvb_{c_k} \gvb_{c_l} &=& - \gvb_{c_l} \gvb_{c_k},\\
 \gv_{c_k} \gvb_{c_l}  &=& - \gvb_{c_l} \gv_{c_k}, \\
 \gv_{c_k} \gv_{c_k}   &=& \phantom{-}\gvb_{c_k} \gvb_{c_k} = 0 \label{eq:zz}.
\end{eqnarray}
In the following, a product of $p$ Grassmann variables (\Gvar) will be called a monomial of degree $p$.
When $p$ is even (odd), such a product will be called an even (odd) monomial. We notice that
an even monomial of \Gvar\ commutes with even and odd monomials of \Gvar. We also remark that exponentials
of pairs of \Gvar\ also commute with even and odd monomials of \Gvar, such an exponential being a sum
of even monomials of \Gvar.
\item \Gvar\ commute with complex numbers and anticommute with Fermionic operators.
\item The fundamental Grassmann calculus rules are
\begin{eqnarray}
\int d\gv_{c_k} \gv_{c_k} = \frac{\partial}{\partial\gv_{c_k}}\gv_{c_k} = 1, \\
\int d\gv_{c_k} = 0.
\end{eqnarray}
\item As a consequence of Eq.~(\ref{eq:zz}), the \Gvar\ $\gv_{c_k}$ play the role of their own 
$\delta$-functions, e.g., for an analytic function $f$ of \Gvar\ (c.f.\ \cite{Zin02}, p.~91), 
we have 
\begin{eqnarray}
\int d\gv_{c_k}\,\gv_{c_k} f(\gv_{c_k}) = f(0).
\label{eq:delta}
\end{eqnarray}
\item The \emph{adjoint} variables $\gvb_{c_k}$ and $\gv_{c_k}$ are independent integration 
variables (c.f.~\cite{Bla85}, p.~28).
\end{itemize}

Concerning Fermionic coherent states, the properties to be used in what follows are:
\begin{itemize}
\item They are eigenstates of second quantized operators
\begin{eqnarray}
\canni_k \, \ket{\vgv_c}=z_{c_k}\, \ket{\vgv_c} \, , \mbox{~~~~~~~~~} \bra{\vgv_c}\, \ccrea_k = \bra{\vgv_c}\, \gvb_{c_k} \, ,
\label{eq:eigenCS}
\end{eqnarray} 
with the eigenvalues being Grassmann variables.
\item They resolve the identity through the closure relation
\begin{equation}
\Id_c=\int{\dnr\vgvb_{c} \, \dnl\vgv_{c} \, \ket{\vgv_c}e^{-\vgvb_c.\vgv_c}\bra{\vgv_c}} \, ,
\label{eq:FCSid}
\end{equation}
where $\Id_c$ means that the resolution of the identity is built for Fock spaces generated by
$\{\vccrea,\vcanni\}$.
We use the short-hand notation $\vgvb_c.\vgv_c\equiv\vgvb_c^t\vgv_c=\sum_{i=1}^n \gvb_{c_i} \gv_{c_i}$.
Finally, $\dnr$ and $\dnl$ represent products of differential elements that are ordered such that
\begin{eqnarray}
\dnr \boldsymbol{x} = dx_n \cdots dx_2 \, dx_1 \, ,\\
\dnl \boldsymbol{x} = dx_1 \, dx_2 \cdots dx_n \, .
\end{eqnarray}
\end{itemize}
As compared to~\cite{Rob09} and many textbooks, we change the 
ordering of the products of differential elements in the integral
in order to anticommute them in a more transparent way, i.e. we use
\begin{equation}
\label{eq:convention:elementaryvariations}
\dnr\vgvb_{c} \, \dnl\vgv_{c},
\end{equation}
which is equivalent to the more commonly used ordering\footnote{
In the commonly used ordering, there is no need to define a particular product ordering 
as $d\gvb_{c_i} d\gv_{c_i}$ are even monomials of \Gvar, and the overall order convention 
is carried only by $d\gvb_{c_i} d\gv_{c_i}$.
}
\begin{equation}
\prod_i d\gvb_{c_i} d\gv_{c_i}.
\end{equation}

\section{Evaluation of the overlaps}
\subsection{Preliminary considerations}
To evaluate the expression for the overlap, we start by inserting two closure relations, 
the left (right) one being based on the single-particle basis of the left (right) 
state, e.g.\ $\bran{\phi_a}\ket{\phi_b} = \bra{\phi_a} \Id_a \Id_b \ket{\phi_b}$
\begin{eqnarray}
\bran{\phi_a}\ket{\phi_b} &=& \int \dnr\vgvb_{a} \dnl\vgv_{a} \dnr\vgvb_b \dnl\vgv_b \nonumber \\
&~& \times \bran{\phi_a}\ket{\vgv_{a}}e^{-\vgvb_a.\vgv_a}\bran{\vgv_{a}}\ket{\vgv_{b}}e^{-\vgvb_b.\vgv_b}\bran{\vgv_{b}}\ket{\phi_b},\nonumber \\
\label{eq:phiaphib}
\end{eqnarray}
where we implicitely use that $\dnr\vgvb_{c}\dnl\vgv_{c}$ ($c=a$, $b$) are even ($2n$) monomials 
of Grassmann differential elements, and thus commuting with Fermionic operators, in order to move 
all differential elements to the very left.

When the single-particle bases of $a$ and $b$ do not span the same subspace of the Hilbert space of single-particle states, 
e.g.\ when they are not linked through a unitary transformation, the resolution of the identity $\Id_a$ ($\Id_b$) works for the
left (right) state alone, and the two closure relations are \emph{not} equivalent. However, there is no loss of generality for 
the following, their non-equivalence being carried by the overlap kernel $\bran{\vgv_{a}}\ket{\vgv_{b}}$.

Given that by definition $\bran{-}\ket{\vgv_{c}}=\bran{\vgv_{c}}\ket{-}=1$, we first 
evaluate the three overlaps
\begin{eqnarray}
\bran{\phi_a}\ket{\vgv_{a}}   &=& \prod^{\tiny \triangleleft}_{i\in\{I^{a}\}} \gv_{a_i} e^{-\frac{1}{2}\sum_{kl}M^{(a)*}_{kl}  \gv^{\,}_{a_k}  \gv^{\,}_{a_l}}, \label{eq:aa} \\
\bran{\vgv_{b}}\ket{\phi_b}   &=& e^{+\frac{1}{2}\sum_{kl}M^{(b)~}_{kl}~ \gvb_{b_k}\gvb_{b_l}} \prod^{\tiny \triangleright}_{j\in\{I^{b}\}} \gvb_{b_j}, \label{eq:bb}\\
\bran{\vgv_{a}}\ket{\vgv_{b}} &=& e^{\sum_{kl} \gvb_{a_k} R_{kl} \gv_{b_l} }. \label{eq:ab}
\end{eqnarray}
The two first expressions use the properties Eq.~(\ref{eq:eigenCS}) of coherent states.
The last one uses the Baker-Campbell-Hausdorff formula,\footnote{The Baker-Campbell-Hausdorff 
formula states that, if $\left[X,\left[X,Y\right]\right]=\left[Y,\left[Y,X\right]\right]=0$,
then $e^X e^Y = e^Y e^X e^{\left[X,Y\right]}$.}
such that
\begin{eqnarray}
\bran{\vgv_{a}}\ket{\vgv_{b}} 	&=& \bra{-}e^{\vgvb_a.\vanni}e^{\vbcrea.\vgv_b}\ket{-}  \nonumber \\
 				&=& \bra{-}e^{\vbcrea.\vgv_b} e^{\vgvb_a.\vanni} e^{\left[\vgvb_a.\vanni,\vbcrea.\vgv_b\right]}\ket{-}.
\end{eqnarray}
Indeed, the latter is applicable because the commutator
\begin{eqnarray}
\left[\vgvb_a.\vanni,\vbcrea.\vgv_b\right]      &=& \sum_{ij} \gvb_{a_i}\left\{\anni_i,\bcrea_j\right\}\gv_{b_j}\\
                                                &=& \sum_{ij} \gvb_{a_i} R_{ij} \gv_{b_j} = \vgvb_a^t R \vgv_b,
\end{eqnarray}
commutes with $\vgvb_a.\vanni$ and $\vbcrea.\vgv_b$, where 
$R_{ij}\equiv\{\anni_i,\bcrea_j\}$ 
denotes the matrix of overlaps of the single-particle states corresponding to $\anni_i$ and $\bcrea_j$.
We finally obtain Eq.~(\ref{eq:ab}) by considering that $\ket{-}$ is a common vacuum for the 
operators $\anni$ and $\banni$.

\subsection{Integration of the reproducing kernel}
\label{sect:reproducing:kernel}

As the next step, we integrate the expression for the reproducing kernel
\begin{eqnarray}
e^{-\vgvb_a.\vgv_a}\bran{\vgv_{a}}\ket{\vgv_{b}}e^{-\vgvb_b.\vgv_b}=e^{-\vgvb_a.\vgv_a+\vgvb_a^t R \vgv_b-\vgvb_b.\vgv_b},\label{eq:reprokernel}
\end{eqnarray}
where we use that exponentials of pairs of \Gvar\ commute, thereby allowing to merge the three 
exponential factors.

Noticing that, in Eq.~(\ref{eq:phiaphib}), $\vgvb_a$ and $\vgv_b$ only appear in the reproducing kernel 
Eq.~(\ref{eq:reprokernel}), we want to integrate these variables separately. In order to do so, we first 
remark that the expression Eq.~(\ref{eq:reprokernel}) can be moved to the very right of Eq.~(\ref{eq:phiaphib}) 
because it commutes with \Gvar. We can as well move the product of differential elements
$\dnr\vgvb_{a}\dnl\vgv_b$ in front of it by virtue of
\begin{eqnarray}
{\ncoord[1.35em,1.25em]{e}\overbrace{\dnr\vgvb_{a}}\dnl\vgv_{a}} \dnr\vgvb_b {\ncoord[0.05em,0.9em]{f}\,} \dnl\vgv_b
&=& (-1)^{2n^2} \dnl\vgv_{a} \dnr\vgvb_b \dnr\vgvb_{a}\dnl\vgv_b \nonumber \\
&=& \dnl\vgv_{a} \dnr\vgvb_b  \underbrace{\dnr\vgvb_{a} \dnl\vgv_b}_{\mbox{\tiny even product}}.
\tikz[overlay,remember picture] { \draw[->] (e) -- ++(0,0.7em) -| (f); }
\end{eqnarray}
With $\dnr\vgvb_{a}\dnl\vgv_b$ being an even product of Grassmann differential elements, it commutes with
$\bran{\phi_a}\ket{\vgv_{a}}$ and $\bran{\vgv_b}\ket{\phi_{b}}$, quantities containing neither the 
variables $\vgvb_{a}$ nor $\vgv_b$, c.f. Eqns.~(\ref{eq:aa}) and~(\ref{eq:bb}). Rewriting Eq.~(\ref{eq:phiaphib}) 
in a more suitable way now gives
\begin{widetext}
\begin{eqnarray}
\bran{\phi_a}\ket{\phi_b} = \int \dnl\vgv_{a} \dnr\vgvb_b \left( \bran{\phi_a}\ket{\vgv_{a}}\bran{\vgv_{b}}\ket{\phi_b}\left(\underbrace{\int \dnr\vgvb_{a}\dnl\vgv_b e^{-\vgvb_a.\vgv_a+\vgvb_a^t R \vgv_b-\vgvb_b.\vgv_b}}_{\mbox{\tiny reproducing kernel integral}}\right)\right) \, .
\label{eq:phiaphib2}
\end{eqnarray}
We now evaluate the reproducing kernel integral.
Provided that $R$ is non-singular,\footnote{
The case of singular $R$ is not equivalent of having zero overlap. As an example, consider
the case of a partially or completely empty single-particle state of the left vacuum which is orthogonal 
to all single-particle states of the right vacuum. In this case, $R$ is singular, whereas the
overlap is not necessarily zero.}$^,$\footnote{
In case of singular $R$, one has to complete the single-particle bases of $\ket{\phi_a}$ 
and $\ket{\phi_b}$ in
order to get an invertible matrix, for example using a Gram-Schmidt orthonormalization procedure.
Still, having a non-singular matrix $R$ is not equivalent to having equivalent bases $a$ and $b$.
} 
we make the change of variables (c.f.\ Ref.~\cite{Zin02} p.~14) 
\begin{eqnarray}
\begin{array}{lcl}
\vgvbe^t&=&\vgvb_a^t-\vgvb_b^t R^{-1} \\
\vgve   &=& \vgv_b-R^{-1}\vgv_a
\end{array}
\end{eqnarray}
such that 
\begin{eqnarray}
\vgvbe^t R \vgve = -\vgvb_a.\vgv_a+\vgvb_a^t R \vgv_b-\vgvb_b.\vgv_b+\vgvb_b^t R^{-1} \vgv_a.
\end{eqnarray}
The reproducing kernel can now be written 
\begin{eqnarray} 
e^{-\vgvb_a.\vgv_a+\vgvb_a^t R \vgv_b-\vgvb_b.\vgv_b} = e^{-\vgvb_b^t R^{-1} \vgv_a} e^{\vgvbe^t R \vgve}.
\end{eqnarray}
The Jacobian of the transformation being one, i.e.\ $\dnr\vgvb_{a} \dnl\vgv_{b} \equiv \dnr\gvbe \dnl\gve$, the integration 
gives
\begin{eqnarray}
&~&\int \dnr\vgvb_a \dnl\vgv_b e^{-\vgvb_a.\vgv_a+\vgvb_{a}^t R \vgv_{b}-\vgvb_b.\vgv_b} \nonumber \\
&~&~~~~~~~= e^{-\vgvb_b^t R^{-1} \vgv_a } \int \dnr\vgvbe \dnl\vgve e^{\vgvbe^t R \vgve} \nonumber \\
&~&~~~~~~~= (-1)^{n}\det\left(R\right) e^{-\vgvb_b^t R^{-1} \vgv_a }
\label{eq:integrepro}
\end{eqnarray}
where we have used the determinant formula outlined in Eq.~(\ref{eq:determinant}) of 
Appendix~\ref{sec:formulae}.

\subsection{Re-expression of the overlap}

Using Eqns.~(\ref{eq:aa}), (\ref{eq:bb}) and (\ref{eq:integrepro}), we are now able to rewrite Eq.~(\ref{eq:phiaphib2}) as
\begin{eqnarray}
\bran{\phi_a}\ket{\phi_b} 	&=& (-1)^{n}\det\left(R\right) \int \dnl\vgv_{a} \dnr\vgvb_b \left( \bran{\phi_a}\ket{\vgv_{a}}\bran{\vgv_{b}}\ket{\phi_b} e^{-\vgvb_b^t R^{-1} \vgv_a }\right)\\
                                &=& (-1)^{n} \det\left(R\right) 
\int \dnl\vgv_{a} \dnr\vgvb_b \left(\prod^{\tiny \triangleleft}_{i\in\{I^{a}\}} \gv_{a_i}\, 
\prod^{\tiny \triangleright}_{j\in\{I^{b}\}} \gvb_{b_j} e^{\left(-\frac{1}{2}\sum_{kl}M^{(a)*}_{kl}  
\gv^{\,}_{a_k}  \gv^{\,}_{a_l} +\frac{1}{2}\sum_{kl}M^{(b)~}_{kl}~ \gvb_{b_k}\gvb_{b_l} -\vgvb_b^t R^{-1} 
\vgv_a \right)}\right)\, , \label{eq:phiaphib3}
\end{eqnarray}
where we used that exponentials of pairs of \Gvar\ commute with \Gvar.
Closely following the notation of Ref.~\cite{Rob09}, we introduce the matrix $\mathbb{M}$ and the vector $\vgvz$
\begin{eqnarray}
\mathbb{M} \equiv \left(\begin{array}{cc} M^{(b)}   & -R^{-1} \\ 
                                     \left(R^{-1}\right)^t  & -M^{(a)*}\end{array}\right), ~~
\vgvz \equiv \left(\begin{array}{c} \vgvb_b \\ 
                                \vgv_a \end{array}\right),
\label{eq:notations}
\end{eqnarray}
such that
\begin{eqnarray}
\frac{1}{2} \vgvz^t  \mathbb{M} \vgvz 	&=& \frac{1}{2}\left(\vgvb_b^t M^{(b)} \vgv_b + \vgv_a^t  R^{-1^t} \vgvb_b - \vgvb_b^t R^{-1} \vgv_a - \vgv_a^tM^{(a)*}\vgv_a\right),\\
					&=& \frac{1}{2}\sum_{ij} \left(M^{(b)}_{ij} \gvb_{b_i}\gvb_{b_j} - M^{(a)*}_{ij}\gv_{a_i}\gv_{a_j}\right) - \sum_{ij}R^{-1}_{ij}\gvb_{b_i}\gv_{a_j}.
\end{eqnarray}
Inserting these definitions into Eq.~(\ref{eq:phiaphib3}), the overlap kernel becomes
\begin{eqnarray}
\bran{\phi_a}\ket{\phi_b} = (-1)^n \det\left(R\right) \int \dnl\vgv_a \dnr\vgvb_b \left(
\prod^{\tiny \triangleleft }_{i\in\{I^{a}\}} \gv_{a_i} 
\prod^{\tiny \triangleright}_{j\in\{I^{b}\}} \gvb_{b_j} \,\, e^{\frac{1}{2} \vgvz^t \mathbb{M} \vgvz }\right).
\label{eq:gv}
\end{eqnarray}

\subsection{Integration over fully-occupied states\label{subsec:intocc}}
We will now integrate variables corresponding to fully occupied states in Eq.~(\ref{eq:gv}) by virtue of 
Eq.~(\ref{eq:delta}). In order to do so, we need to move the corresponding differential elements to the 
very right, which gives a sign factor because of the anticommutation of Grassmann differential elements. 
For example, such a rearrangement for a single variable in $\dnl \vgv_{c}$ and $\dnr \vgvb_{c}$ gives
\begin{eqnarray}
\dnl \vgv_{c}  = d\gv_{c_1} \cdots \ncoord[1.08em,1.3em]{a}\overbrace{d\gv_{c_k}} 
\cdots d\gv_{c_n} \,\ncoord[0,0.6em]{b}\, = \left(\prod^{\tiny \triangleleft}_{i\ne k}d\gv_{c_i}\right) (-1)^{n-k} d\gv_{c_k},
\tikz[overlay,remember picture] {\draw[->] (a) -- ++(0,0.3em) -| (b);}
\end{eqnarray}
\begin{eqnarray}
\dnr \vgvb_{c} = d\gvb_{c_n} \cdots \ncoord[1.08em,1.3em]{c}\overbrace{d\gvb_{c_k}} 
\cdots d\gvb_{c_1}  \,\ncoord[0,0.6em]{d}\, = \left(\prod^{\tiny \triangleright}_{i\ne k}d\gvb_{c_i}\right)(-1)^{k-1}d\gvb_{c_k}.
\tikz[overlay,remember picture] {\draw[->] (c) -- ++(0,0.3em) -| (d);}
\end{eqnarray}
When there is more than one fully occupied state, the repeated application of this procedure gives
\begin{eqnarray}
\dnl \vgv_{c} &=& \left(\prod^{\tiny \triangleleft}_{i\notin\{I^{c}\}}d\gv_{c_i}\right)   
	\left(\prod^{\tiny \triangleright}_{k\in\{I^{c}\}}(-1)^{n-k}d\gv_{c_{k}}\right),\label{eq:ben1}\\
\dnr \vgvb_{c}&=& \left(\prod^{\tiny \triangleright}_{i\notin\{I^{c}\}}d\gvb_{c_i}\right) 
	\left(\prod^{\tiny \triangleleft}_{k\in\{I^{c}\}}(-1)^{k-1}d\gvb_{c_{k}}\right),\label{eq:ben2}
\end{eqnarray}
where in each equation we notice the opposite order for the products over indices corresponding to fully 
occupied states as compared to other states. Combining Eqns.~(\ref{eq:ben1}) and~(\ref{eq:ben2}) gives
\begin{eqnarray}
\dnl\vgv_a \dnr\vgvb_b 	&=& 
	\left(\prod^{\tiny \triangleleft}_{i\notin\{I^{a}\}}d\gv_{a_i}\right) 
	\ncoord[5.14em,2.6em]{x}\overbrace{\left(\prod^{\tiny \triangleright}_{k\in\{I^{a}\}}(-1)^{n-k}d\gv_{a_{k}}\right)} 
	\left(\prod^{\tiny \triangleright}_{j\notin\{I^{b}\}}d\gvb_{b_j}\right) 
	\left(\prod^{\tiny \triangleleft}_{l\in\{I^{b}\}}(-1)^{l-1}d\gvb_{b_{l}}\right) \, \ncoord[0,0.6em]{y}\,\label{eq:bennotyet1}
	\tikz[overlay,remember picture] {\draw[->] (x) -- ++(0,0.3em) -| (y);}\\
	 		&=& (-1)^{nr_a}\left(\prod^{\tiny \triangleleft}_{i\notin\{I^{a}\}}d\gv_{a_i}\right) 
	\left(\prod^{\tiny \triangleright}_{j\notin\{I^{b}\}}d\gvb_{b_j}\right) 
	\left(\prod^{\tiny \triangleleft}_{l\in\{I^{b}\}}(-1)^{l-1}d\gvb_{b_{l}}\right)
	\left(\prod^{\tiny \triangleright}_{k\in\{I^{a}\}}(-1)^{n-k}d\gv_{a_{k}}\right).\label{eq:bennotyet2}
\end{eqnarray}
The additional sign in Eq.~(\ref{eq:bennotyet2}) comes from the commutation of variables $k\in\{I^{a}\}$ to the very right,
as indicated in Eq.~(\ref{eq:bennotyet1}). Defining the sign factor 
\begin{equation}
\sigma \equiv (-1)^{\sum_{k\in\{I^{a}\}}(n+k)+\sum_{k\in\{I^{b}\}}k} \, , 
\end{equation}
Eq.~(\ref{eq:bennotyet2}) finally gives, after a suitable rearrangement of the sign factors
\begin{eqnarray}
\dnl\vgv_a \dnr\vgvb_b = 
(-1)^{nr_a+r_b}
\sigma
\left(\prod^{\tiny \triangleleft }_{i\notin\{I^{a}\}}d\gv_{a_i} \right) 
\left(\prod^{\tiny \triangleright}_{j\notin\{I^{b}\}}d\gvb_{b_j}\right) 
\left(\prod^{\tiny \triangleleft }_{l\in\{I^{b}\}   }d\gvb_{b_l}\right)
\left(\prod^{\tiny \triangleright}_{k\in\{I^{a}\}   }d\gv_{a_k} \right).
\end{eqnarray}
The ordering of the differential elements corresponding to indices of fully occupied states are now in the
appropriate order with respect to their associated products in Eq.~(\ref{eq:gv}) to perform their integration.

By virtue of Eq.~(\ref{eq:delta}), the integration over fully occupied levels of the integrand in Eq.~(\ref{eq:gv}) gives
\begin{eqnarray}
\int \left(\prod^{\tiny \triangleleft}_{k\in\{I^{b}\}}d\gvb_{b_{k}}\right)\left(\prod^{\tiny \triangleright}_{k\in\{I^{a}\}}d\gv_{a_{k}}\right)
\left( \prod^{\tiny \triangleleft }_{i\in\{I^{a}\}} \gv_{a_i} 
\prod^{\tiny \triangleright}_{j\in\{I^{b}\}} \gvb_{b_j} \,\, e^{\frac{1}{2} \vgvz^t \mathbb{M} \vgvz }\right) = \left(e^{\frac{1}{2} \vgvz^t \mathbb{M} \vgvz} \right)_{\gvz_i=0\,\, \forall i\in\{I^{b}\}, n-i\in\{I^{a}\}} \, .
\end{eqnarray}
With this, Eq.~(\ref{eq:gv}) can be rewritten as
\begin{eqnarray}
\bran{\phi_a}\ket{\phi_b} = (-1)^n (-1)^{nr_a+r_b}\sigma \det\left(R\right) 
\int \left(\prod^{\tiny \triangleleft}_{i\notin\{I^{a}\}}d\gv_{a_i}\right) \left(\prod^{\tiny \triangleright}_{j\notin\{I^{b}\}}d\gvb_{b_j}\right) 
\left(e^{\frac{1}{2} \vgvz^t \mathbb{M} \vgvz} \right)_{\gvz_i=0\,\, \forall i\in\{I^{b}\}, n-i\in\{I^{a}\}},
\label{eq:gvbis}
\end{eqnarray}
where the integration only runs over variables of indices associated to not fully-occupied states.

%
%
\subsection{Integration over the remaining variables}
 
We now define a new matrix $\mathbb{M}_{r}$, sub-matrix of $\mathbb{M}$ where rows and columns of
indices $i \in \{I^{b}\}$ of fully occupied states in $\ket{\phi_b}$ and of incices $j+n$, where $j\in \{I^{a}\}$ 
are fully occupied levels in $\ket{\phi_a}$, have been removed. 
The corresponding appropriate vector $\vgvz_r$ is built from $\vgvz$ in the same
manner, removing components with indices  $i \in \{I^{b}\}$ and $j+n$ such that $j\in \{I^{a}\}$. 
The matrix $\mathbb{M}_{r}$ is skew-symmetric with dimension $N_r\times N_r$, and the vector $\vgvz_r$ has 
$N_r$ elements, with $N_r=2n-(r_a+r_b)$.

For an example where there are two fully occupied states in each quasiparticle vacuum, with indices 
$i$, $k$ for the state $\ket{\phi_b}$ and indices $j$, $l$ for the state $\ket{\phi_a}$, respectively, 
the matrix $\mathbb{M}_{r}$ and the vector $\vgvz_r$ can be schematically represented as
\begin{eqnarray}
~\\
~\\
\begin{array}{rclcrclc}
  \mathbb{M}_r &=&~~ \left(\begin{array}{cccccc}
\ncoord[-1.25em,-0.15em]{hbb1d}
\ncoord[-1.25em, 0.40em]{hbb2d}
\ncoord[-0.15em, 1.55em]{vbb1d}
\ncoord[ 0.50em, 1.55em]{vbb2d}
                           & \phantom{-\left(M^{(a)}\right)^*} & 
\ncoord[ 0.5em, 0.5em]{vm1}
&
\ncoord[ 0.15em, 1.55em]{vaa1d}
\ncoord[ 1.35em, 1.55em]{vaa2d}
&         &
\ncoord[ 0.55em,-0.15em]{hbb1f}
\ncoord[ 0.55em, 0.40em]{hbb2f}
 \\
                           & M^{(b)} & & & -R^{-1} & \\
\ncoord[-0.5em,-0.35em]{hm1}
                           &         & & &         & 
\ncoord[ 0.5em,-0.35em]{hm2}
\\
\ncoord[-1.25em,-0.3em]{haa1d}
\ncoord[-1.25em, 0.7em]{haa2d}
                           &         & & &         & 
\ncoord[0.55em,-0.3em]{haa1f}
\ncoord[0.55em, 0.7em]{haa2f}
\\
                           & \left(R^{-1}\right)^t & & & -\left(M^{(a)}\right)^* & \\
\ncoord[-0.15em, -0.35em]{vbb1f}
\ncoord[ 0.50em, -0.35em]{vbb2f}
                           &         & 
\ncoord[ 0.5em, -0.35em]{vm2}
&
\ncoord[ 0.15em,-0.35em]{vaa1f}
\ncoord[ 1.35em,-0.35em]{vaa2f}
&         & 
                         \end{array}\right)
& \mbox{~~~} & \vgvz_r & = &~~ \left(\begin{array}{c}
                             \phantom{\left(l^b\right)^t}
				\ncoord[-2.5em,-0.15em]{hb1d}\ncoord[0.75em,-0.15em]{hb1f}
				\ncoord[-2.5em,-0.40em]{hb2d}\ncoord[0.75em,-0.40em]{hb2f}\\
                             \vgvb_b\\
                             \phantom{l^b}\ncoord[-1.55em, -0.35em]{hv1} \ncoord[0.75em, -0.35em]{hv2}\\
                             \phantom{l^b}\\
                             \vgv_a\\
                             \phantom{\left(l^b\right)^t}
                             \end{array}\right)
\end{array}
\tikz[overlay,remember picture] {\draw[color=red]  (vbb1d) node[above]{i} -- (vbb1f);}
\tikz[overlay,remember picture] {\draw[color=red]  (vbb2d) node[above]{k} -- (vbb2f);}
\tikz[overlay,remember picture] {\draw[color=red]  (hbb1d) node[left]{k}  -- (hbb1f);}
\tikz[overlay,remember picture] {\draw[color=red]  (hbb2d) node[left]{i}  -- (hbb2f);}
\tikz[remember picture,overlay] {\path (hbb1d)++(20.4em,0.0em) coordinate (hb1d);}
\tikz[remember picture,overlay] {\path  (hb1d)++(3.9em,0.0em)  coordinate (hb1f);}
\tikz[remember picture,overlay] {\path (hbb2d)++(20.4em,0.0em) coordinate (hb2d);}
\tikz[remember picture,overlay] {\path  (hb2d)++(3.9em,0.0em)  coordinate (hb2f);}
\tikz[overlay,remember picture] {\draw[color=red]  (hb1d)  node[left]{k}  -- (hb1f); }
\tikz[overlay,remember picture] {\draw[color=red]  (hb2d)  node[left]{i}  -- (hb2f); }
\tikz[overlay,remember picture] {\draw[color=blue] (haa1d) node[left]{l}  -- (haa1f);}
\tikz[overlay,remember picture] {\draw[color=blue] (haa2d) node[left]{j}  -- (haa2f);}
\tikz[overlay,remember picture] {\draw[color=blue] (vaa1d) node[above]{j} -- (vaa1f);}
\tikz[overlay,remember picture] {\draw[color=blue] (vaa2d) node[above]{l} -- (vaa2f);}
\tikz[remember picture,overlay] {\path (haa1d)++(20.4em,0.0em) coordinate (ha1d);}
\tikz[remember picture,overlay] {\path  (ha1d)++(3.9em,0.0em)  coordinate (ha1f);}
\tikz[remember picture,overlay] {\path (haa2d)++(20.4em,0.0em) coordinate (ha2d);}
\tikz[remember picture,overlay] {\path  (ha2d)++(3.9em,0.0em)  coordinate (ha2f);}
\tikz[overlay,remember picture] {\draw[color=blue]  (ha1d)  node[left]{l}  -- (ha1f); }
\tikz[overlay,remember picture] {\draw[color=blue]  (ha2d)  node[left]{j}  -- (ha2f); }
\tikz[overlay,remember picture] {\draw[dotted]       (hm1)                -- (hm2)  ;}
\tikz[overlay,remember picture] {\draw[dotted]       (vm1)                -- (vm2)  ;}
\tikz[remember picture,overlay] {\path (hm1)++(20.4em,0.0em) coordinate (hv1);}
\tikz[remember picture,overlay] {\path (hv1)++(3.0em,0.0em) coordinate (hv2);}
\tikz[overlay,remember picture] {\draw[dotted]       (hv1)                -- (hv2)  ;}
\label{eq:redset}
\end{eqnarray} 
where labeled rows and columns have been removed from the original objects $\mathbb{M}$ and $\vgvz$.
Using the submatrix $\mathbb{M}_{r}$ and subvector $\vgvz_r$, Eq.~(\ref{eq:gvbis}) can be rewritten as
\begin{eqnarray}
\bran{\phi_a}\ket{\phi_b} = (-1)^n (-1)^{nr_a+r_b}\sigma \det\left(R\right) 
\int \left(\prod^{\tiny \triangleleft}_{i\notin\{I^{a}\}}d\gv_{a_i}\right) \left(\prod^{\tiny \triangleright}_{j\notin\{I^{b}\}}d\gvb_{b_j}\right) 
\left(e^{\frac{1}{2} \vgvz_r^t \mathbb{M}_r \vgvz_r} \right) \, . 
\label{eq:gvter}
\end{eqnarray}
In order to apply the Pfaffian formula Eq.~(\ref{eq:pfaffian}), the order of 
$\prod^{\tiny \triangleleft}_{i\notin\{I^{a}\}}d\gv_{a_i}$ has to be reversed, such that the 
differential elements in Eq.~(\ref{eq:gvter}) are in the appropriate order with respect to the 
matrix $\mathbb{M}_r$. This is achieved by reversing the order of $\prod^{\tiny \triangleleft}_{i\notin\{I^{a}\}}d\gv_{a_i}$ 
\begin{eqnarray}
\prod^{\tiny \triangleleft}_{i\notin\{I^{a}\}}d\gv_{a_i} = (-1)^{(n-r_a)(n-r_a-1)/2}\prod^{\tiny \triangleright}_{i\notin\{I^{a}\}}d\gv_{a_i}.
\label{eq:pfaffreord}
\end{eqnarray}
The sign factor from the reversal of the product $\gv_1\cdots\gv_n=(-1)^{n(n-1)/2} \; \gv_n\cdots\gv_1$ 
can be obtained by induction. The differential elements are now in the appropriate order
\begin{eqnarray}
\mathrm{d}^{N_r}_{\tiny\,\, \triangleright\,} \vgvz_r 	
		&=& \prod^{\tiny \triangleright}_{i\notin\{I^{a}\}}d\gv_{a_i} \prod^{\tiny \triangleright}_{j\notin\{I^{b}\}}d\gvb_{b_j} \\
		&=& (-1)^{(n-r_a)(n-r_a-1)/2} \prod^{\tiny \triangleleft}_{i\notin\{I^{a}\}}d\gv_{a_i} \prod^{\tiny \triangleright}_{j\notin\{I^{b}\}}d\gvb_{b_j} \, ,
\end{eqnarray}
such that we can now integrate the remaining variables using the Pfaffian formula, Eq.~(\ref{eq:pfaffian}),
\begin{eqnarray}
\bran{\phi_a}\ket{\phi_b} &=& (-1)^n (-1)^{nr_a+r_b}\sigma \det\left(R\right) (-1)^{(n-r_a)(n-r_a-1)/2}
\int \mathrm{d}^{N_r}_{\tiny\,\, \triangleright\,}\vgvz_r\, e^{\frac{1}{2}\vgvz_r^t \mathbb{M}_{r}\vgvz_r}\label{eq:benalmostdone}\\
&=& (-1)^{n(n+1)/2}(-1)^{r_a(r_a-1)/2}(-1)^{r_a+r_b}\sigma \det\left(R\right) \mbox{pf}\left(\mathbb{M}_{r}\right).
\label{eq:gvqua}
\end{eqnarray}

\end{widetext}
This formula, however, assumes $N_r$ to be even, see appendix~\ref{sec:formulae}. 
When $N_r$ is odd, $\ket{\phi_a}$ and $\ket{\phi_b}$ have in fact different \emph{number parity}~\cite{Rin80}. 
In that case, the overlap $\bran{\phi_a}\ket{\phi_b}$ is automatically zero. From the definition of the 
Pfaffian, which by definition is zero for skew-symmetric matrices of odd rank, we can thus notice that formula
Eq.~(\ref{eq:gvter}) can still be applied. In particular, $(-1)^{r_a+r_b}$ will always be one except when multiplied
by zero, allowing to drop this sign factor in the final formula. We thus summarize the final expression 
for the overlap as
\begin{eqnarray}
\label{eq:result}
\bran{\phi_a}\ket{\phi_b} 
= s_n s_{r_a} \sigma \det\left(R\right) \mbox{pf}\left(\mathbb{M}_{r}\right) \, , 
\end{eqnarray}
where
\begin{eqnarray}
s_n &=& (-1)^{n\left(n+1\right)/2} \, , \\
s_{r_a} &=& (-1)^{r_a(r_a-1)/2} \, , \\
\sigma&=&(-1)^{\sum_{k=1..r_a}(n+i_{k_a})+\sum_{k=1..r_b}i_{k_b}} \, .
\end{eqnarray}
The sign factors depend on the number of states in the single-particle bases $n$, the number of fully occupied 
states $r_a$ in $a$, and the indices $i_{k_c}$ of fully occupied states in the bases $c=a$, $b$.

Equation~(\ref{eq:result}) provides the generalization of Eq.~(7) of Ref.~\cite{Rob09} to the overlap
between different quasiparticle vacua with an arbitrary number of fully occupied single-particle states. 
In particular, it can be applied to overlaps that involve an odd number of blocked quasiparticle states, 
a case not considered at all in Refs.~\cite{Rob09,Rob11}, and for the special case of symmetry restoration 
only in Ref.~\cite{Ber11x}.
Moreover, when an even number of particles is fully occupied (either for blocked $2n$ quasiparticle 
states, or as a result of the minimization, or both), Eq.~(\ref{eq:result}) provides a formally justified
alternative to the regularization of the matrix $\mathbb{M}$ performed in~\cite{Rob11}.

Besides this important generalization, there is another noteworthy difference to previous work 
by Robledo~\cite{Rob09,Rob11}. Indeed, Eq.~(\ref{eq:result}) is directly expressed in the
single-particle bases of $\ket{\phi_a}$ and $\ket{\phi_b}$, respectively, that allow for the most 
compact representation of these quasiparticle vacua. 
In particular, Eq.~(\ref{eq:result}) can also be applied without invoking a complete 
single-particle basis spanning the single particle subspace $a\,\cup\,b$, cf.\ the discussions above.
However, as explained there, if the matrix $R$ is singular, one is forced to complete each single-particle 
basis until $\mbox{det}\left(R\right)\ne 0$ is achieved. 

For quasiparticle vacua for which there are no fully occupied states in their respective
single-particle basis, it is easy to show that Eq.~(7) of Ref.~\cite{Rob09} is recovered.

\section{Some Illustrative Examples}

\begin{figure}[!t]
\includegraphics {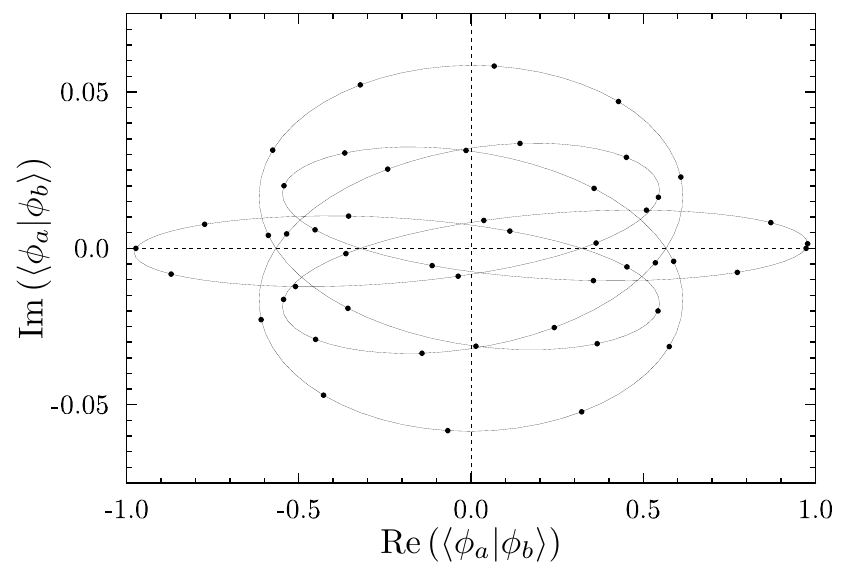}
\caption{\label{fig:mg25_qp:2}
Real and imaginary parts of the overlap without particle-number projection for the 
lowest one-quasiparticle state in \nuc{25}{Mg} obtained with SIII. 
The Euler angles $\alpha$ and $\beta$ are held fixed at values of $\alpha = 1.25^\circ$ 
and $\beta = 7.17^\circ$, wheras $\gamma$ is varied in the interval
$[0,720^\circ]$ with a discretization of 288 points. Filled circles on the curve
represent a discretization of 48 points in the interval $[0,720^\circ]$, 
which is sufficient to converge observables. Note the difference in scale of 
real and imaginary parts.
}
\end{figure}

The determination of the phase of the overlap by the widely used techniques that 
rely on a Taylor expansion of the overlap around a matrix element of known phase 
\cite{Har82,Ena99,Val00} works very well when restricting the calculations 
to time-reversal invariant HFB states. However, it becomes increasingly 
difficult when time-reversal is broken, as it happens for odd-$A$ or 
odd-odd nuclei, or for states obtained with cranked HFB, cf.\ the examples
discussed in Refs.~\cite{Har82,Ena99,Oi05}.

\begin{figure}[!t]
\includegraphics {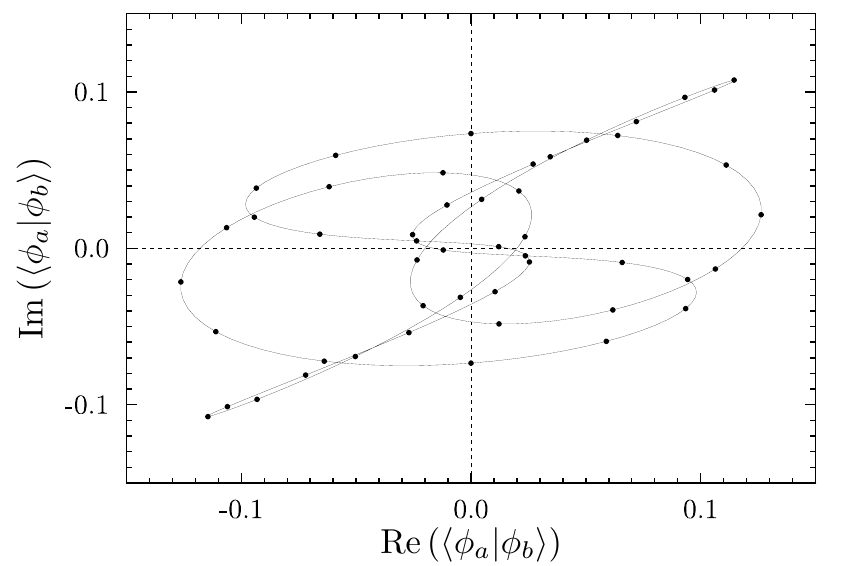}
\caption{\label{fig:mg25_qp}
Same as Fig.~\ref{fig:mg25_qp:2}, but for $\alpha = 43.75^\circ$ and $\beta = 71.94^\circ$. 
}
\end{figure}

We have implemented Eq.~(\ref{eq:result}) into our numerical codes for 
particle-number and angular-momentum restored GCM calculations based
on triaxial HFB states \cite{Ben09a,Ben11x}, using the routines for the 
calculation of the Pfaffian of Ref.~\cite{Gon10}.
We now present three examples where techniques to follow the overlap
through Taylor expansion might fail and the direct calculation of the
overlap becomes a necessity. 
These illustrations will show trajectories in the complex plane of 
overlaps of quasi-particle vacua as obtained during angular-momentum 
projection
\begin{eqnarray}
\bran{\phi_a}\ket{\phi_b} = \bra{\varphi}\hat{R}\left(\alpha,\beta,\gamma\right)\ket{\varphi},
\end{eqnarray}
where $\hat{R}$ is the rotation operator that depends on the three Euler angles
$\alpha$, $\beta$, and $\gamma$.

The first two examples are presented in Figs.~\ref{fig:mg25_qp:2}
and~\ref{fig:mg25_qp}. They illustrate the trajectory of the overlap in the 
complex plane when varying the Euler angles $\gamma$ when projecting
the lowest self-consistent one-quasiparticle state of \nuc{25}{Mg}, 
for two different combinations of $\alpha$ and $\beta$.
The first one, Fig.~\ref{fig:mg25_qp:2}, illustrates that real and 
imaginary parts of the overlap can vary on quite different scales. 
In this particular case, most of the modulus of the overlap is 
carried by the real part, and the phase of the overlap is most
of the time either close to zero or close to $\pm\pi$. 
Unless the discretization of the Euler angles is carefully adapted,
the phase of the overlap might change by almost $\pi$ when crossing the 
imaginary axis, which is very difficult to distinguish from a discontinuity
by $\pi$ encountered when having lost the phase.
The second example, Fig.~\ref{fig:mg25_qp},
obtained for a different combination of Euler angles 
$\alpha$ and $\beta$, shows that the trajectory of the overlap in the 
complex plane may exhibit cusps, which might again be difficult to 
resolve when discretizing Euler angles.

\begin{figure}[!t]
\includegraphics{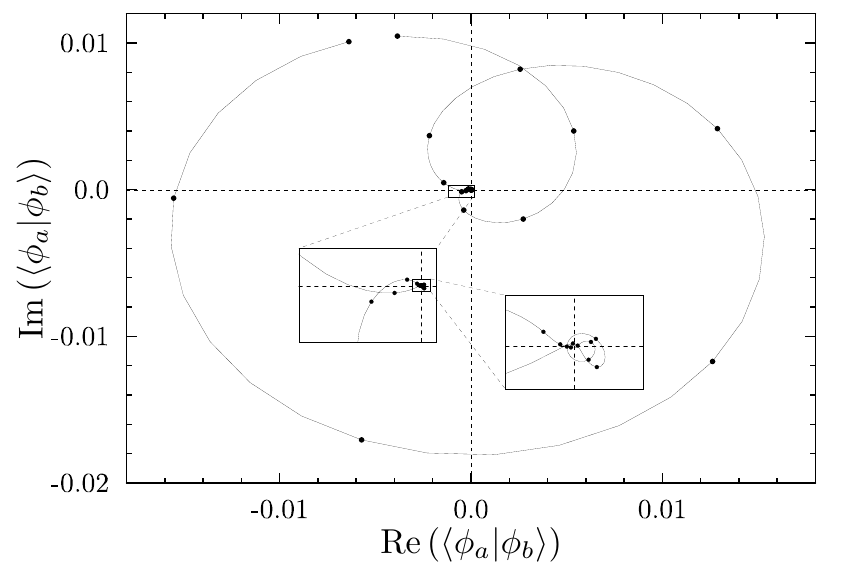}
\caption{\label{fig:mg24_j8}
Real and imaginary parts of the overlap without particle-number projection
for the cranked $I = 8 \hbar$ state in \nuc{24}{Mg} obtained with SIII. The 
Euler angles $\alpha$ and $\beta$ are held fixed at values of 
$\alpha = 23.75^\circ$ and $\beta= 66.96^\circ$, respectively, wheras $\gamma$ 
is varied in the interval $[0,360^\circ]$ with a discretization of 144 points. 
The inserts amplify the zone at very small overlaps, whereas filled circles on 
the curves represent a discretization of 24 points in the interval $[0,360^\circ]$, 
which is sufficient to converge observables.
}
\end{figure}

In Fig.~\ref{fig:mg24_j8}, we present the trajectory of the overlap of a
high-spin state in \nuc{24}{Mg}, obtained from cranked 
HFB+Lipkin Nogami \cite{Ter95}.
The two inserts illustrate that variations may occur on very different 
scales with quite involved structures.

These three examples demonstrate that a Taylor-expansion-based 
algorithm to determine unambiguously the sign of the overlap may 
become difficult in applications that use time-reversal invariance
breaking quasiparticle vacua.
Indeed, it should be able to resolve discontinuities or 
cusps, or other involved structures. The latter might happen at very 
different scales, and the discretization must be chosen in order to
account for all these details. Furthermore, we expect the complexity 
of such trajectories to increase with increasing intrinsic angular 
momentum. A direct determination of the overlap is thus a considerable 
improvement not only from a formal point of view, but also from the 
perspective of the complexity of reliable algorithms for the computation 
of the overlap on the one hand and of computing time on the other hand, 
as it will often allow for the use of a smaller number of combinations 
of Euler angles in angular-momentum projection.

\section{Discussion and Outlook}

To summarize our main findings
\begin{enumerate}
\item
An extension of the Thouless parameterization of
quasiparticle vacua with completely filled single-particle states allows to
calculate the overlap directly in a formalism based on Grassmann 
algebra and coherent states similar to the one outlined in Ref.~\cite{Rob09}, 
but in such a manner that fully occupied or empty single-particle levels 
are automatically taken care of without any need for the manipulation 
(or regularization) of matrices elaborated in Ref.~\cite{Rob11}.
\item
This extension of the Thouless expression allows to handle all 
possible quasiparticle vacua that have completely filled states, 
i.e.\ also 1-, 2-, \ldots $n$-quasiparticle states, not just 
quasiparticle vacua that can be expressed as limits of 
fully-paired quasiparticle vacua as in Ref.~\cite{Rob11}. 
\item
The handling of blocked states is not restricted to pure symmetry restoration
as the one proposed in Ref.~\cite{Ber11x}, and therefore can be also applied 
when the non-rotated left and right states are different, which is necessary 
for GCM calculations.
\item
Our final expression for the overlap allows for the calculation of the
overlap of two quasiparticle vacua that are expressed in two 
different single-particle bases that do not span the same sub-space
of the Hilbert space of single-particle states. 
The knowledge of a complete basis spanning both single-particle bases
is not needed, as compared to Ref.~\cite{Rob11}.
In this way, the technique can be directly implemented in codes that 
use a coordinate space representation of the quasiparticle vacua in 
terms of their canonical single-particle bases~\cite{Bon90,Val00,Ben09a,Ben11x}.
\end{enumerate}
The expression has been implemented into our numerical codes for 
particle-number and angular-momentum restored GCM calculations based
on triaxial HFB states using the full space of occupied single-particle 
states \cite{Ben09a,Ben11x}. It has been extensively tested for 
symmetry restoration and for the calculation of non-diagonal
matrix elements for symmetry-restored GCM calculations without 
encountering cases where it fails. 

By contrast, the technique ``to follow the phase'' from Ref.~\cite{Val00,Har82}, 
or the one to follow the overlap in the complex plane from Ref.~\cite{Ena99} 
require often to use a very fine discretization to resolve the sign ambiguity 
when projecting on angular momentum as soon as time-reversal invariance of the
HFB states is broken. In particular, it may become necessary to use a 
discretization of the integrals over Euler or gauge angles that are
much finer than what is actually needed to converge observables.
In addition, the direct calculation of the overlap also has  the advantage
to avoid complicated coding for the set-up of a reliable Taylor-expansion-based
algorithm, in particular since reliable routines to compute pfaffians are
available~\cite{Gon10}. Finally, in the general case of configuration mixing 
where there might not be a symmetry that establishes a reference sign for the 
overlap, such a direct calculation of the overlap could become to be mandatory.
 
In summary, we report an expression for the overlap between arbitrary 
quasiparticle vacua that is easy to calculate and that is very robust 
in realistic applications. It is a key ingredient for the extension
of symmetry restoration and Generator Coordinate Method-type calculations
to angular-momentum-optimized states, either by cranking, or by blocking.

\begin{acknowledgments}
This work was supported by the Agence Nationale de la Recherche 
under Grant No.~ANR 2010 BLANC 0407 "NESQ". 
The computations were performed using HPC resources from 
GENCI-IDRIS (Grant 2011-050707).
\end{acknowledgments}

\begin{appendix}

\section{Some remarkable Gaussian integrals over Grassmann variables \label{sec:formulae}}

For the readers' convenience, we give the identities that represent determinants and pfaffians 
as integrals over Grassmann variables using our convention for the order of the differential 
elements, Eq.~(\ref{eq:convention:elementaryvariations}).

\subsection{Determinant}
\label{sect:determinant}
The first one is the determinant identity that is defined, for a given matrix $M$, as 
\begin{eqnarray}
\det \left(M\right) 	&=& \int \prod_i \left(d\gv_i d\gvb_i\right) \exp\left(\sum_{i,j=1}^n \gvb_i M_{ij} \gv_j\right)\\
			&=& (-1)^n \int \dnr\vgvb \dnl \vgv \exp\left(\sum_{i,j=1}^n \gvb_i M_{ij} \gv_j\right) \, ,
\label{eq:determinant}
\end{eqnarray}
where the first equation is the expression from Ref.~\cite{Zin02}, p.~13, Eq.~(1.67), and 
the second uses an alternative convention in the ordering of differential elements.
The latter differs by a sign because of the anticommutation of $\mathcal{G}$-variables.

\subsection{Pfaffian}

The second one is the Pfaffian identity that is defined for a skew-symmetric matrix $A$ of 
dimension $2n\times 2n$,  as (see \cite{Zin02}, p.15, Eq.~(1.80)):
\begin{eqnarray}
\mbox{pf}\left(A\right) &=& \int d\gv_{2n}\cdots d\gv_{2}d\gv_{1} \exp\left(\frac{1}{2}\sum_{ij=1}^{2n} \gv_i A_{ij} \gv_j\right)\\
                        &=& \int \mathrm{d}^{2n}_{\,\tiny \triangleright}\vgv \exp\left(\frac{1}{2}\sum_{ij=1}^{2n} \gv_i A_{ij} \gv_j\right)
\, .
\label{eq:pfaffian}
\end{eqnarray}
To obtain the correct sign, it is crucial that the differential elements are in 
the same order as the indices of the matrix $A_{ij}$. This implies sometimes to make 
some manipulations to bring the entire expression into the proper form, as for example 
between Eq.~(\ref{eq:gvter}) and Eq.~(\ref{eq:benalmostdone}).

As can be easily seen, this definition is only valid for matrices $A$ of \emph{even} rank.
The determinant and the Pfaffian of a skew-symmetric matrix $A$ are related by
\begin{equation}
\left[\mbox{pf} \left(A\right)\right]^2 = \det \left(A\right)\, .
\end{equation}
As the determinant of a skew-symmetric matrix of \emph{odd} rank is always zero, the pfaffian
of such a matrix is defined to be zero as well.
\end{appendix}

\end{document}